\documentclass[smallextended]{svjour3}       
\smartqed  
\usepackage{graphicx}
\usepackage{multirow}%
\usepackage{amsmath,amssymb,amsfonts}%
\usepackage{mathrsfs}%
\usepackage[title]{appendix}%
\usepackage{xcolor}%
\usepackage{textcomp}%
\usepackage{manyfoot}%
\usepackage{booktabs}%
\usepackage{algorithm}%
\usepackage{algorithmicx}%
\usepackage{algpseudocode}%
\usepackage{listings}%
\usepackage[misc]{ifsym}

\newtheorem{obs}{Observation}
\usepackage{orcidlink}
\usepackage{natbib}
\usepackage{todonotes}
\newcommand{\opt}{\mathrm{OPT}}
\newcommand{\area}{\mathrm{AREA}}
\newcommand{\afdh}{\mathrm{AFDH}}
\newcommand{\nfdh}{\mathrm{NFDH}}
\newcommand{\ffds}{\mathrm{FFDS}}

\newcommand{\lb}{\mathrm{LB}}
\newcommand{\ptasvar}{\varepsilon}

\newcommand{\areaoflargeitem}{\ptasvar^{2p}}
\newcommand{\sizeofsmallitem}{\ptasvar^{p + 3}}
\newcommand{\areaofsmallitem}{\ptasvar^{2p + 6}}
\newcommand{\sizeofgroups}{\ptasvar^{2p+1}}

\newcommand{\minsum}{Min-Sum Bin Packing Problem}
\newcommand{\msbpp}{MSBPP}
\newcommand{\squareminsum}{Square \minsum}
\newcommand{\smsbpp}{S\msbpp}
\newcommand{\minsumweighted}{Mininum Weighted Sum Bin Packing Problem}
\newcommand{\mwsbpp}{MWSBPP}

\journalname{Journal of Combinatorial Optimization}
\begin{document}

\title{Approximation algorithms for the {\squareminsum}
}


\author{Rachel Vanucchi Saraiva\orcidlink{0000-0001-6197-603X}         \and
        Rafael C. S. Schouery\orcidlink{0000-0002-0472-4810} 
}


\institute{\Letter R. V. Saraiva \at
			  Universidade Estadual de Campinas (UNICAMP), Instituto de Computação, Brazil \\
              \email{r185961@dac.unicamp.br}           
           \and
           R. C. S. Schouery \at
		   Universidade Estadual de Campinas (UNICAMP), Instituto de Computação, Brazil
}

\date{Received: date / Accepted: date}

\maketitle
\begin{abstract}
In this work, we study the {\squareminsum} (\smsbpp), where a list of~$n$ square items has to be packed into square bins of dimensions $1 \times 1$ with no overlap between the areas of the items. The bins are indexed (starting at one) and the cost of packing each item is equal to the index of the bin in which it is placed. The objective is to minimize the total cost of packing all items, which is equivalent to minimizing the average cost of items. The problem has applications in minimizing the average time of logistic operations such as cutting stock and delivery of products. We prove that classic algorithms for two-dimensional bin packing that order items in non-increasing order of size, such as Next Fit Decreasing Height or Any Fit Decreasing Height heuristics, can have an arbitrarily bad performance for SMSBPP\@. We, then, present an algorithm with an approximation ratio of $\sqrt{205}-12+\delta$ ($\approx 2.3178+\delta$), for any $\delta > 0$, and running time $O(n \log n)$. Finally, we also present a PTAS for the problem.
\end{abstract}

\section{Introduction}\label{section_intro}
The Bin Packing Problem (BPP) is a classic NP-hard problem where a list of items with varying sizes must be packed into bins of identical capacity. It models relevant logistic problems for industries such as the storage of products and digital files, loading of cargo into vehicles, task scheduling on several machines, and many others (see \cite{applications}).

In the classic BPP, the goal is to minimize the number of bins needed to pack all items, which, in real applications, leads to minimizing storage costs and waste of resources. \cite{epstein2007} defined the {\minsum} ({\msbpp}), where bins are indexed and the cost of packing an item is equal to the index of the bin in which it is placed. The objective is to minimize the total cost of packing all items, which is equivalent to minimizing the average cost of the items. For example, if each bin represents the items a single vehicle carries for one trip, then the solution of MSBPP minimizes the average number of trips that it takes to transport an item.

The same authors have also defined a weighted version of the problem, the {\minsumweighted} ({\mwsbpp}), where each item has an associated weight to it, and the cost of packing an item is the index of the bin in which it is placed multiplied by the weight of the item. This represents the idea that certain items might have priority over others and should be packed first. 

The {\msbpp} is a special case of a batch-scheduling problem where a single machine can process, in parallel, a batch of jobs of limited total size, and the objective is to minimize the sum of completion times of all jobs, with the batches being processed sequentially. If every job (and therefore every batch) has a processing time of~$1$, the completion time of jobs in the $j$-th batch is~$j$, and thus the problem is equivalent to {\msbpp}. This problem is noted as $1|p\text{-\emph{batch}},s_j,p_j=1|\sum C_j$ in the three-field notation for scheduling problems introduced by \cite{batchnotation}. In the same way, the {\mwsbpp} is a particular case of a similar problem, where the objective is to minimize the weighted sum of completion times. For more details on batch scheduling, we refer to the recent survey done by~\cite{batchsurvey}.

Both {\msbpp} and {\mwsbpp} define items and bins as one-dimensional, that is, each item is associated with a single size, and the sum of the sizes of all items packed in a bin cannot exceed the capacity of the bin. In this paper, we study the Square Min-Sum Bin Packing Problem (SMSBPP), where the items and bins are squares, and items must be packed within the internal area of the bin, with no overlap between items. The complete formal definition of the problem is given in Section~\ref{section_def}. The definition of items as two-dimensional captures extra challenges not represented by the one-dimensional definition when it comes to the placement of physical items in a bin.

In the following section, we discuss previous works related to min-sum and square bin packing. In Section~\ref{section_def}, we formally define the problem and the symbols used throughout the paper. We also show how classic heuristics for two-dimensional bin packing do not have constant approximation ratios when applied to SMSBPP, motivating the design of new algorithms for this problem. Then, in Section~\ref{section_approx}, we present, for any~$\delta > 0$, an algorithm for SMSBPP with an approximation ratio of $\sqrt{205}-12+\delta$ ($\approx 2.3178+\delta$) and running time $O(n \log n)$. In Section~\ref{section_ptas}, we present a PTAS for the problem. Finally, in Section~\ref{section_concl}, we discuss our conclusions from the study and present directions for future work.

\section{Previous work\label{section_prem}}
To the best of our knowledge, there is no previous work on the {\minsum} with items and bins with more than one dimension. For {\mwsbpp}, \cite{epstein2007} present a PTAS and a Dual-PTAS\@. As for {\msbpp}, \cite{epstein2018} present a Next Fit Increasing heuristic with an absolute approximation ratio of at most~$2$ and an asymptotic approximation ratio of at most~$1.6188$, along with a heuristic with an asymptotic approximation ratio of at most~$1.5604$. The authors also present a PTAS for {\msbpp} similar to the one for {\mwsbpp}\@.

Many works exist in relation to classic BPP with multidimensional items and bins. We refer to the survey done by \cite{approxsurvey} for further details on approximation algorithms for the problem.

For packing two-dimensional items, we will focus on level-oriented heuristics, which pack items into levels within the bin. A level is a strip with width equal to the width of the bin and height equal to the first item packed in the level. Usually, the items are also ordered by non-increasing order of height. 

Among these heuristics, there is Next Fit Decreasing Height (NFDH), where items are packed in the current level until an item cannot fit in, which leads to a new level being opened. The previous levels do not receive any additional items. Similarly, a new bin is only opened once it cannot fit a new level, and, once it is opened, no levels are created in the previous bins as well. 

Other level-oriented heuristics are those from the family Any Fit Decreasing Height (AFDH), where a new level is only opened if the current item cannot fit in any of the previous levels, and, similarly, a new bin is only opened when a level will not fit in any of the previous bins.

For NFDH, \cite{ffdh} show that if~$\varepsilon$ is an upper bound on the areas of the items packed by the algorithm, then the unoccupied area left in every bin except possibly the last one is at most~$2\varepsilon$. \cite{square2} has also proved that, if the items packed by NFDH are all squares of size at most~$1/3$, then the area occupied in every bin, except possibly the last one, is at least~$9/16$. We use NFDH and these results in our proposed algorithms. However, we also prove in the following section that NFDH and AFDH heuristics alone do not have a constant approximation ratio for SMSBPP.

Still on the topic of algorithms for classic BPP, \cite{2dinapprox} have shown that there is no APTAS for the general case of BPP with rectangular items. They have, however, proved the existence of an APTAS for the packing of $d$-dimensional cubes into unit $d$-dimensional bins. 

When it comes to the special case of packing only square items, \cite{squarepacking} have shown that NFDH and AFDH heuristics can fit a set of items of total area at most~$1/2$ into a single bin. \cite{ffds} have presented the First Fit Decreasing Size (FFDS) algorithm for optimal packing of items of size greater than~$1/3$. The way the algorithm works is described as follows. It orders the items of size greater than~$1/2$ in non-decreasing order of size and packs them one per bin, in the bottom left corner of the bin. Then, it orders the remaining items in non-increasing order of size and checks if the first item in the list fits in any of the bins already opened. If it fits, then the first three items of the list are packed into the first bin they fit, one in each unoccupied corner of the bin. If the item does not fit any existing bin, then a new bin is opened and the first four items of the list are packed in it. The process then repeats until all items have been packed. Figure~\ref{fig_ffds} shows an example of a packing by FFDS\@.

\begin{figure}[H]
\centering
\begin{tikzpicture}[scale=0.5]
\draw (0, 0) rectangle (4, 4);
\draw (5, 0) rectangle (9, 4);
\draw (10, 0) rectangle (14, 4);
\filldraw[draw=black, fill=lightgray] (0, 0) rectangle (2.3, 2.3);
\filldraw[draw=black, fill=lightgray] (5, 0) rectangle (7.5, 2.5);
\filldraw[draw=black, fill=gray] (2.5, 0) rectangle (4, 1.5);
\filldraw[draw=black, fill=gray] (0, 2.5) rectangle (1.5, 4);
\filldraw[draw=black, fill=gray] (2.5, 2.5) rectangle (4, 4);

\filldraw[draw=black, fill=gray] (10, 0) rectangle (12, 2);
\filldraw[draw=black, fill=gray] (12.1, 0) rectangle (14, 1.9);
\filldraw[draw=black, fill=gray] (10, 2.1) rectangle (11.9, 4);
\filldraw[draw=black, fill=gray] (12.2, 2.2) rectangle (14, 4);
\end{tikzpicture}
\caption[Example of packing done by FFDS.]{Example of packing done by FFDS}\label{fig_ffds}
\end{figure}

Note that at most four items of size greater than~$1/3$ can be packed in a single bin, and FFDS packs as many bins with this number of items as it is possible. Therefore, if we reorder the bins of the FFDS packing in non-increasing order of the number of items, the solution is also optimal for an instance of SMSBPP with these items. Combining NFDH and FFDS, \cite{square2} presents a~$2$-approximation for BPP with square items. We use a similar approach in our approximation for SMSBPP in Section~\ref{section_approx}.

\section{Definitions and preliminaries}\label{section_def}

We define the Square Min-Sum Bin Packing Problem (SMSBPP) as follows: the input is a list~$I$ of~$n$ square items. We define the size of a square as the size of its height (and thus also its width), and, for~$i$ such that $1 \leq i \leq n$, the size of the~{$i$-th} square in the list is~$s_i \in (0,1]$. All items in the list must be packed in indexed square bins~$B_1, \dots, B_m$ of side~$1$ (where $m$ is arbitrary), and the cost of packing an item in~$B_j$ is~$j$. The objective is to minimize the total cost ${\sum_{j=1}^m j \cdot |B_j|}$ of packing all items, in which~$|B_j|$ is the number of items packed in bin~$B_j$. A feasible packing of items in a bin is one where all items are within the internal area of the bin, with no intersection between the internal area of an item and the borders of the bin or the internal area of another item. Items are packed with their sides parallel to the sides of the bin.

We now show that the NFDH and AFDH heuristics mentioned in the previous section are not approximation algorithms for the SMSBPP\@, as ordering items by non-increasing order of size is unhelpful for this problem. 

\begin{lemma}\label{lemma_inapprox}
NFDH and AFDH heuristics do not have a constant approximation ratio for SMSBPP\@.
\end{lemma}
\begin{proof}
Let $t \geq 3$ be an integer. We create an instance~$I$ of SMSBPP with~$t^2$ items of size~$1/t$, which we refer to as small items from here on, and~$t$ items of size ${1-1/t}$, which we refer to as large items. Any AFDH heuristic will first pack one large item per bin, since two large items cannot fit together in the same bin. Each bin can then still fit $t+1$ small items, one in the same level as the large item, and~$t$ in a new level above it. Therefore, $t$ bins are used to pack all the items, $t-1$ of them containing~$t+2$ items, and the last one containing~$2$ items. Figure~\ref{fig_ffdh_minsum} shows an example of this packing. The value of the solution is, therefore,
\begin{equation*}
\afdh(I) = \sum_{j=1}^{t-1} j \cdot (t+2) + 2t = \frac{t^3 + t^2 + 2t}{2}.
\end{equation*}

\begin{figure}[H]
	\centering
	\begin{tikzpicture}[scale=0.5]
	\draw (0, 0) rectangle (3, 3);
	\draw (4, 0) rectangle (7, 3);
	\draw (8, 0) rectangle (11, 3);
	
	\draw[draw=black, fill=lightgray] (0, 0) rectangle (2,2);
	\draw[draw=black, fill=lightgray] (4, 0) rectangle (6,2);
	\draw[draw=black, fill=lightgray] (8, 0) rectangle (10,2);
	
	\draw[draw=black, fill=lightgray] (2, 0) rectangle (3,1);
	\draw[draw=black, fill=lightgray] (6, 0) rectangle (7,1);
	\draw[draw=black, fill=lightgray] (10, 0) rectangle (11,1);
	
	\draw[draw=black, fill=lightgray] (0, 2) rectangle (3,3);
	\draw[draw=black, fill=lightgray] (4, 2) rectangle (7,3);
	
	\foreach \i in {1,...,2}{
		\draw (1*\i, 2) -- (1*\i, 3);
		\draw (4+1*\i, 2) -- (4+1*\i, 3);
	}
	\draw[draw=white] (0,-1) -- (1, -1);
	\end{tikzpicture}
	\begin{tikzpicture}[scale=0.5]
	\draw[draw=black, fill=gray] (0, 0) rectangle (3, 3);
	\draw (4, 0) rectangle (7, 3);
	\draw (8, 0) rectangle (11, 3);
	\draw (12, 0) rectangle (15, 3);
	
	\draw[draw=black, fill=gray] (4, 0) rectangle (6, 2);
	\draw[draw=black, fill=gray] (8, 0) rectangle (10, 2);
	\draw[draw=black, fill=gray] (12, 0) rectangle (14, 2);
	
	\draw (0,1) -- (3,1);
	\draw (0,2) -- (3,2);
	\draw (1,0) -- (1,3);
	\draw (2,0) -- (2,3);
	\end{tikzpicture}
	\caption[Packing done by an AFDH heuristic compared to another possible packing.] {Packing of~$I$ with $t=3$ done by an AFDH heuristic (above) compared to another possible packing (below)}\label{fig_ffdh_minsum}
\end{figure}

Now consider another solution, also depicted in Figure~\ref{fig_ffdh_minsum}, where all small items are packed in the first bin, and then $t$ bins are opened to pack one large item each. The value of this solution is
\begin{equation*}
t^2 + \sum_{j=2}^{t+1} j = t^2 + \frac{(t+3)t}{2} = \frac{3t^2 + 3t}{2} \geq \opt(I),
\end{equation*}
in which $\opt(I)$ is the value of an optimal solution for~$I$.

Therefore, the approximation ratio of any AFDH heuristic is at least
\begin{equation*}
\frac{\afdh(I)}{\opt(I)} \geq \frac{t^3 + t^2 + 2t}{2} \cdot \frac{2}{3t^2 + 3t} \geq \frac{t}{3}.
\end{equation*}

The NFDH heuristic also first packs one large item per bin. Then, only the last of these bins packs $t+1$ small items, and the remaining $t^2-t-1$ small items are packed in a new bin, as shown in Figure~\ref{fig_nfdh_minsum}. Thus, the value of the solution is 
\begin{equation*}
\nfdh(I) = \sum_{j=1}^{t-1} j + t(t+2) + (t+1)(t^2-t-1)= \frac{2t^3 + 3t^2 - t - 2}{2}.
\end{equation*}
\begin{figure}[t]
\centering
\begin{tikzpicture}[scale=0.5]
\draw (0, 0) rectangle (3, 3);
\draw (4, 0) rectangle (7, 3);
\draw (8, 0) rectangle (11, 3);
\draw (12, 0) rectangle (15, 3);

\draw[draw=black, fill=gray] (0, 0) rectangle (2, 2);
\draw[draw=black, fill=gray] (4, 0) rectangle (6, 2);
\draw[draw=black, fill=gray] (8, 0) rectangle (10, 2);

\draw[draw=black, fill=gray] (8, 2) rectangle (11,3);
\draw[draw=black, fill=gray] (10, 0) rectangle (11,1);

\draw[draw=black, fill=gray] (12, 1) rectangle (14,2);
\draw[draw=black, fill=gray] (12, 0) rectangle (15,1);

\draw (9, 2) -- (9,3);
\draw (10, 2) -- (10,3);

\draw (13, 0) -- (13,2);
\draw (14, 0) -- (14,2);
\draw (12, 2) -- (15,2);
\end{tikzpicture}
\caption[Packing done by NFDH.] {Packing of~$I$ with $t=3$ done by NFDH}\label{fig_nfdh_minsum}
\end{figure}

Since $2t^2 > t + 2$ for every $t \geq 2$, the approximation ratio of NFDH is at least
\begin{equation*}
\frac{\nfdh(I)}{\opt(I)} \geq \frac{2t^3 + 3t^2 - t - 2}{3t^2 + 3t} > \frac{2t^3 + t^2}{3t^2 + 3t} > \frac{t}{3},
\end{equation*}
and the lemma follows.\qed
\end{proof}

\section{A $(\sqrt{205}-12+\delta)$-approximation\label{section_approx}}

We now present, for any~$\delta > 0$, a $(\sqrt{205}-12+\delta)$-approximation for SMSBPP. The algorithm is also a $2$-approximation for the special case where all items are of size at most~$1/3$.

The algorithm is described as follows. For any~$\delta > 0$, we define a constant $n_0(\delta)$ (the exact definition is given at the end of the section, along with the pseudocode for the algorithm). If $n < n_0(\delta)$, the algorithm solves the problem optimally by enumerating all possible partitions of items (as there are at most $B_{n_0}$ partitions, where $B_{n_0}$ is the $n_0$-th Bell number) and testing each partition to see if it is a valid packing. While this enumeration can be done in constant time when $n < n_0(\delta)$, it can still be very time-consuming. Thus, we show that~$n_0(\delta)$ can be chosen to be fairly small, something that will inevitably lead to a worse approximation ratio, as $\delta$ will be large.

If $n \geq n_0(\delta)$, first we classify items according to their size. Items of size at most~$1/3$ are \emph{small items}, items of size greater than~$1/2$ are \emph{large items}, and the remaining are \emph{medium items}. We also define~$\area(C)$ as the sum of the areas of items in a set~$C$. We order the items in non-decreasing order of their sizes and then partition them into groups $G_1, \dots, G_q$ such that each group~$G_i$ receives items until $\area(G_i) > 1$, or until there are no more items in the instance (for the final group).

Let~$r$ be the number of groups containing only small items, and let~$G^S_{r+1}$ be the set of small items in~$G_{r+1}$. We pack the items of each group $G_1, \dots, G_r$ and~$G^S_{r+1}$, in order, with NFDH\@. Afterward, we pack the large and medium items with FFDS and sort the bins in the solution found by FFDS in a non-increasing order of the number of items. Finally, we place the bins of the packing found with FFDS for medium and large items after the packing found by NFDH for small items. As both NFDH and FFDS can be implemented with time complexity~$O(n \log n)$~(see \cite{ffds}), the complexity of our algorithm is also~$O(n \log n)$.

\subsection{Bounds}

Let~$A(I)$ be the value of the solution given by our algorithm for an instance~$I$ with $n \geq n_0(\delta)$. For analysis of the approximation ratio, we now find lower bounds~$\lb_1$ and~$\lb_2$ on~$\opt(I)$ such that ${A(I)\leq \alpha \lb_1 + \lb_2}$ for some ${\alpha > 0}$, as this implies that the algorithm is a $(\alpha+1)$-approximation.

The first bound, $\lb_1 = \sum_{i=1}^q i|G_i|$, is obtained from the definition of the groups~$G_i$. As \cite{epstein2018} have shown, if $\area(G_i) > 1$ for all $i$ in $\{1, \dots, t\}$, and groups $G_1, \dots, G_t$ contain the smallest items of the instance, then no feasible solution can pack more than ${\sum_{i=1}^t} |G_i|$ items in the first~$t$ bins, and so the following lemma applies.

\begin{lemma}[Epstein et al., 2018]\label{lemma_approx_bound}
$\opt(I) \geq \sum_{i=1}^q i|G_i|$.
\end{lemma}

For ease of notation, let $R = \sum_{i=1}^r i|G_i| + (r+1)|G^S_{r+1}|$. As mentioned before, \cite{square2} has shown that every bin with only small items packed by NFDH except possibly the last one has occupied area of at least~$9/16$. Since the area of any small item is at most~$1/9$ and~$\area(G_i)$ is greater than~$1$ by a single overfilling item, $\area(G_i) \leq 10/9 < 9/8$ for $i \leq r$, and ${\area(G^S_{r+1}) \leq 1}$. Therefore, for $i \leq r+1$, the small items of the first~$i$ groups are packed by NFDH in at most~$2i$ bins, and the cost of a small item of~$G_i$ in this packing is at most~$2i$. Thus, the total cost of packing all small items by our algorithm is at most
\begin{equation*}
\sum_{i=1}^r 2i|G_i| + (2r+2)|G^S_{r+1}| = 2R.
\end{equation*} 

By Lemma~\ref{lemma_approx_bound}, this means the algorithm is a $2$-approximation for instances with only small items.

The lower bound given by Lemma~\ref{lemma_approx_bound} is, however, less helpful for the analysis of instances with large items. In fact, since a group~$G_i$ can contain up to four large items, and yet only a single large item can be packed in each bin, their cost might be close to~$4i$ in our packing. Because of that, we define another lower bound on~$\opt(I)$ based on the fact that the packing of FFDS is optimal for the subset of large and medium items of the instance.

Let~$\ffds(d)$ be the cost of packing large and medium items with FFDS (and bin reordering) after a bin of index~$d$ (if $d=0$, then the bins of FFDS are the first bins in the solution). We make the following observation.

\begin{obs}\label{obs_subsets}
If $A$ and $B$ are disjoint subsets of items in an instance, then $\opt(A \cup B) \geq \opt(A) + \opt(B)$.
\end{obs}
\begin{proof}
Let $S^*$ be an optimal solution for $A \cup B$. Note that $S^*$ induces a feasible solution for~$A$, where the items of the subset are packed in the same bins they are in $S^*$. Similarly, it also induces a feasible solution for~$B$. The sum of the value of these feasible solutions equals $\opt(A \cup B)$, and therefore $\opt(A \cup B)$ is an upper bound of the value of $\opt(A) + \opt(B)$.\qed
\end{proof}

By Lemma~\ref{lemma_approx_bound},~$R$ is a lower bound of the cost of an optimal solution for the small items. The packing done by FFDS is optimal for the large and medium items. Therefore, by Observation~\ref{obs_subsets}, another lower bound, $\lb_2$, on the value of an optimal solution for the whole instance is
\begin{equation*}
\lb_2 = R + \ffds(0).
\end{equation*}

We now develop bound~$\lb_1$ a little further based on the FFDS packing. Let~$k$ be the number of medium items in the instance plus the number of large items packed in the same bin as some medium item by FFDS, and let~$b$ be the number of large items packed alone in a bin by FFDS\@. Since~$\area(G_i)$ is greater than~$1$ by a single item, any group~$G_i$ contains at most~$9$ medium items and at most~$4$ large items. Therefore, the number of groups containing medium items plus the large items packed with a medium item by FFDS is at least~$\lfloor k/9 \rfloor$, and the number of groups containing the large items packed alone by FFDS is at least~$\lfloor b/4 \rfloor$. Thus, since $\lfloor k/9 \rfloor \geq k/9 - 1$ and $\lfloor b/4 \rfloor \geq b/4 - 1$,
\begin{align*}
\lb_1 &= \sum_{i=1}^q i|G_i| \\
	&\geq R + \sum_{i=1}^{\lfloor k/9 \rfloor} 9(r+i) + \sum_{i=1}^{\lfloor b/4 \rfloor}4(r+\lfloor k/9 \rfloor + i)\\
	&\geq R + 9r\left(\frac{k}{9}-1\right) + \frac{9\left(\frac{k}{9}-1\right)\frac{k}{9}}{2}+ 4\left(r+\frac{k}{9}-1\right)\left(\frac{b}{4}-1\right) \\
	&\qquad+ \frac{4\left(\frac{b}{4}-1\right)\frac{b}{4}}{2}\\
	&=R + rk -13r + \frac{k^2}{18} - \frac{17k}{18} +rb + \frac{kb}{9} - \frac{3b}{2} + 4 + \frac{b^2}{8}.
\end{align*}

For ease of notation, let $B = rk -13r + \frac{k^2}{18} - \frac{17k}{18} +rb + \frac{kb}{9} - \frac{3b}{2} + 4 + \frac{b^2}{8}$. With that, $\lb_1 \geq R + B$.

\subsection{Analysis of the algorithm}

We now begin to compare the value of the solution of our algorithm with the bounds on~$\opt(I)$ we defined. We assume there is at least one medium or large item in the instance, as Lemma~\ref{lemma_approx_bound} already shows that the algorithm is a $2$-approximation otherwise. Then, we make the following observation.

\begin{obs}\label{obs_ffds_compare}
$\ffds(d) = (k+b)d + \ffds(0)$.
\end{obs}
\begin{proof}
The number of items packed by FFDS is $k+b$, and placing the solution after~$d$ bins raises the cost of each item by~$d$ compared to a solution with only the bins of FFDS\@.\qed
\end{proof}

We will use this observation to compare the value of the solution of our algorithm with the~$\lb_2$ bound. For that, we need a bound on the value of~$A(I)$. As mentioned before, the cost of packing the small items with the algorithm is at most~$2R$. Then the large and medium items are packed by FFDS with bin reordering. Since at most $2r+2$ bins are used to pack the small items,
\begin{equation*}
A(I) \leq 2R + \ffds(2r+2).
\end{equation*}

For ease of notation, let $\Delta_\alpha= A(I) - \alpha \lb_1 - \lb_2$. We want to prove that $\Delta_\alpha \leq 0$ for some value of~$\alpha$. By Observation~\ref{obs_ffds_compare}, we have that 
\begin{equation*}
A(I) - \lb_2 \leq R + (2r+2)(k+b).
\end{equation*}

Let~$s$ be the number of small items in the instance, that is,
\begin{equation*}
s = \sum_{i=1}^r |G_i| + |G_{r+1}^S|.
\end{equation*} 

As one can see, $\Delta_\alpha$ depends not only on $\alpha$, but also on $r$, $s$, $k$, and $b$. It also depends on $n$, as $n = s+k+b$. Thus, in order to bound $\Delta_\alpha$, we will first remove the dependence on $s$, $k$, and $b$, finding an upper bound on $\Delta_\alpha$ that only depends on $\alpha$, $r$, and $n$. With this in mind, we have the following lemma.

\begin{lemma}\label{lemma_approx_bk}
If $\alpha \geq 1$, $\Delta_\alpha \leq \frac{18(r+1)^2}{\alpha} -13r^2 + 22r +35 -\alpha\left(\frac{r^2}{2} -\frac{r}{2} -\frac{113}{8} + n\right)$.
\end{lemma}
\begin{proof}
By the definition of~$\Delta_\alpha$ and the fact that~$\lb_1 \geq R + B$, we have that
\begin{equation*}
\Delta_\alpha = A(I) - \alpha \lb_1 - \lb_2 \leq R + (2r+2)(k+b) - \alpha(R + B).
\end{equation*}

Recalling now that $R = \sum_{i=1}^r i|G_i| + (r+1)|G^S_{r+1}|$, we obtain
\begin{align*}
\Delta_\alpha &\leq (2r+2)(k+b) -(\alpha-1)\left(\sum_{i=1}^r i|G_i| + (r+1)|G^S_{r+1}|\right) -\alpha B\\
	&\quad -\alpha\left(s - \sum_{i=1}^r |G_i| - |G_{r+1}^S|\right).
\end{align*}

For ${i \leq r}$, items in group~$G_i$ have area of at most~$1/9$. Therefore, since $\area(G_i) > 1$, that means~$|G_i| \geq 10$, and as $\alpha \geq 1$, we have that
\begin{align*}
\Delta_\alpha &\leq (2r+2)(k+b) -(\alpha-1)\left(\sum_{i=1}^r 10i\right) -\alpha B -\alpha\left(s - \sum_{i=1}^r 10\right)\\
		&=(2r+2)(k+b) - (\alpha-1)(5r^2+5r)- \alpha(B+s-10r)\\
		&=(2r+2)(k+b) +5r^2+5r -\alpha(5r^2-5r+B+s).
\end{align*}

Recall that $s = n - k - b$. With that,
\begin{equation*}
\Delta_\alpha \leq (2r+2)(k+b) +5r^2+5r -\alpha(5r^2-5r+B+n-k-b),
\end{equation*}
and, since $B = rk -13r + \frac{k^2}{18} - \frac{17k}{18} +rb + \frac{kb}{9} - \frac{3b}{2} + 4 + \frac{b^2}{8}$, we have that
\begin{align*}
\Delta_\alpha &\leq (2r+2)(k+b) + 5r^2 + 5r\\
	&\quad-\alpha\left(5r^2 + rk -18r + \frac{k^2}{18} -\frac{35k}{18} + rb + \frac{kb}{9} - \frac{5b}{2} + 4 + \frac{b^2}{8} + n\right).
\end{align*}

Let $g_b$ and $g_k$ be the derivatives of the right-hand side expression (the upper bound on $\Delta_\alpha$) above with respect to~$b$ and~$k$, respectively. We have that
\begin{equation*}
	g_b(r,k,b)=2r+2-\alpha\left(r+\frac{k}{9}-\frac{5}{2}+\frac{b}{4}\right)
\end{equation*} and 
\begin{equation*}
	g_k(r,k,b)=2r+2-\alpha\left(r+\frac{k}{9}-\frac{35}{18}+\frac{b}{9}\right).
\end{equation*} 

We have a critical point $(b, k)$ of the function when $g_b = g_k = 0$, that is,
\begin{equation*}
-\frac{5}{2}+\frac{b}{4} = -\frac{35}{18}+\frac{b}{9}
\end{equation*}
from where we conclude that $b = 4$ and, consequently,
\begin{equation*}
k = 9\left(\frac{2r+2}{\alpha} -r + \frac{3}{2}\right).
\end{equation*}

Now let $h_{bb}$, $h_{kk}$, and~$h_{bk}$ be the partial second derivatives of the expression, that is, $h_{bb}=-\frac{\alpha}{4}$, $h_{kk} = -\frac{\alpha}{9}$, and $h_{bk} = -\frac{\alpha}{9}$. By the second derivative test, we have that $h_{bb}h_{kk}-h^2_{bk} > 0$ and $h_{bb} < 0$, and thus the critical point found is a maximum point.

Therefore, we can substitute the value of~$b$ at this maximum point in our lower bound on~$\Delta_\alpha$, and with that
\begin{align*}
\Delta_\alpha &\leq (2r+2)(k+4)+ 5r^2 + 5r\\
	&\quad- \alpha\left(5r^2 + rk -18r + \frac{k^2}{18}-\frac{35k}{18} + 4r + \frac{4k}{9} - \frac{20}{2} + 4 + \frac{4^2}{8} + n\right)\\
	&= \left(2r+2-\alpha\left(r-\frac{3}{2}\right)\right)k - \frac{\alpha k^2}{18}+ 13r + 8 -\alpha\left(5r^2 - 14r -4 +n \right)\\
	&\quad +5r^2.
\end{align*}

Next we can substitute the value of~$k$ at this maximum point in the lower bound as well. Note first that the first term of the expression is
\begin{equation*}
\left(2r+2-\alpha\left(r-\frac{3}{2}\right)\right)k = \alpha\left(\frac{2r+2}{\alpha} -r + \frac{3}{2}\right)k= \frac{\alpha k^2}{9},
\end{equation*}
and with that,
\begin{align*}
\Delta_\alpha &\leq  \frac{\alpha k^2}{9} - \frac{\alpha k^2}{18}+13r + 8-\alpha\left(5r^2 - 14r -4 +n \right)+5r^2\\
	&= \frac{\alpha k^2}{18} + 13r + 8 -\alpha\left(5r^2 - 14r -4 +n \right)+5r^2\\
	&=\frac{18(r+1)^2}{\alpha} - 18r^2 + 9r + 27 +\frac{9\alpha r^2}{2}-\frac{27\alpha r}{2} + \frac{81\alpha}{8} +13r + 8 \\
	&\quad-\alpha\left(5r^2 - 14r -4 +n \right)+5r^2\\
	&=\frac{18(r+1)^2}{\alpha} -13r^2 + 22r +35-\alpha\left(\frac{r^2}{2} -\frac{r}{2} -\frac{113}{8} + n\right),
\end{align*}
and the lemma follows.\qed
\end{proof}

Lemma~\ref{lemma_approx_bk} gives us an upper bound on~$\Delta_\alpha$ that does not depend on the values of~$s$, $b$, and~$k$. Similarly, we now find another bound that does not depend on the value of~$r$ either.

\begin{lemma}\label{lemma_approx_alpha}
If $\alpha > \sqrt{205}-13$, then $\Delta_\alpha \leq \frac{-2718-1653\alpha-57\alpha^2}{2(72-52\alpha-2\alpha^2)}\alpha- \alpha n$.
\end{lemma}
\begin{proof}
By Lemma~\ref{lemma_approx_bk},
\begin{align*}
\Delta_\alpha &\leq \frac{18(r+1)^2}{\alpha} -13r^2 + 22r +35-\alpha\left(\frac{r^2}{2} -\frac{r}{2} -\frac{113}{8} + n\right)\\
&\leq \left(\frac{36-26\alpha-\alpha^2}{2\alpha}\right)r^2+\left(\frac{72 + 44\alpha + \alpha^2}{2\alpha} \right)r+\frac{18}{\alpha} +35 + \frac{113\alpha}{8} - \alpha n.
\end{align*}

Let $g'_r$ and $g''_r$ be the first and second derivatives (with respect to $r$) of the right-hand side expression, respectively. That is, $g'_r = \frac{36}{\alpha}(r+1)-26r +22 -\alpha r +\frac{\alpha}{2}$, and $g''_r = \frac{36}{\alpha}-26 -\alpha$. Since $\alpha > \sqrt{205}-13$, we have $g''_r \leq 0$. Thus, the bound is maximized when $g'_r = 0$. The value for~$r$ at this maximum point is
\[
\frac{36}{\alpha}(r+1)-26r +22 -\alpha r +\frac{\alpha}{2} = 0
\]
and, we conclude that,
\[
r = \frac{-72-44\alpha-\alpha^2}{72-52\alpha-2\alpha^2}.
\]

Thus, by substituting this value on our lower bound, we get
\begin{align*}
\Delta_\alpha &\leq \left(\frac{36-26\alpha-\alpha^2}{2\alpha}\right)r^2 + \left(\frac{72 + 44\alpha +\alpha^2}{2\alpha}\right)r+\frac{18}{\alpha} +35 +\frac{113\alpha}{8} - \alpha n\\
&=\left(\frac{72+44\alpha+\alpha^2}{4\alpha}\right)r+\frac{18}{\alpha} +35 +\frac{113\alpha}{8} - \alpha n\\
&=\frac{-(72+44\alpha+\alpha^2)^2}{4\alpha(72-52\alpha-2\alpha^2)}+\frac{18}{\alpha} +35 +\frac{113\alpha}{8} - \alpha n\\
&=\alpha\left(\frac{-2718-1653\alpha-57\alpha^2}{2(72-52\alpha-2\alpha^2)}- n\right),
\end{align*}
and the lemma follows.\qed
\end{proof}

We can now finally prove the approximation ratio of the algorithm.

\begin{theorem}\label{thm_approx}
The algorithm proposed is a $\sqrt{205}-12+\delta$-approximation for any chosen $\delta > 0$.
\end{theorem}

\begin{proof}
Recall that $\Delta_\alpha= A(I) - \alpha \lb_1 - \lb_2$, and thus $\Delta_\alpha \leq 0$ implies that the algorithm is an $(\alpha+1)$-approximation. Lemma~\ref{lemma_approx_alpha} shows that the value of~$\Delta_\alpha$ decreases linearly as~$n$ increases. In particular, we have that $\Delta_\alpha \leq 0$ when
\begin{equation*}
n \geq \frac{-2718-1653\alpha-57\alpha^2}{2(72-52\alpha-2\alpha^2)}.
\end{equation*}

Therefore, for every~$\delta > 0$, we can define $\alpha = \sqrt{205}-13+\delta$ and
\begin{equation*}
n_0(\delta) = \left\lceil\frac{-2718-1653\alpha-57\alpha^2}{2(72-52\alpha-2\alpha^2)}\right\rceil
\end{equation*}
such that the algorithm is an $(\alpha+1)$-approximation.\qed
\end{proof}

For examples on the implications of Theorem~\ref{thm_approx} according to the choice of~$\alpha$: choosing $\alpha = 1.4$ and plugging the value in the expression above shows that the algorithm is a $2.4$-approximation when $n_0(\delta) = 545$, while choosing $\alpha = 2$ shows that the algorithm is a $3$-approximation when $n_0(\delta) = 79$. The limit of $\frac{-2718-1653\alpha-57\alpha^2}{2(72-52\alpha-2\alpha^2)}$ as $\alpha$ tends to infinity is $57/4 = 14.25$. Thus, we have that $n_0(\delta) \geq 15$. 

Now that we have the definition of $n_0(\delta)$, the full description of the algorithm is as follows.

\begin{algorithm}\label{alg_approx}
	\caption{$(\sqrt{205}-12+\delta)$-approximation for SMSBPP}
	\begin{algorithmic} 
		\Require {List $I$ of $n$ square items, $\delta > 0$}
		\State Step 1: Compute $n_0(\delta) \gets \left\lceil\frac{-2718-1653\alpha-57\alpha^2}{2(72-52\alpha-2\alpha^2)}\right\rceil$,  where $\alpha = \sqrt{205}-13+\delta$
		\If {$n < n_0(\delta)$}
		Solve the problem by enumerating all possible partitions.
		\Else 
		\State Step 2: Order items in non-decreasing order of size
		\State Step 3: Divide items into groups $G_1, \dots, G_q$ such that each group~$G_i$ receives items until $\area(G_i) > 1$, or there are no more items in the instance
		\State Step 4: Pack the small items of each group, in order, with NFDH
		\State Step 5: Pack the medium and large items with FFDS
		\State Step 6: Sort the bins obtained in Step 5 by non-increasing order of the number of items, and place them after the bins obtained in Step 4
		\EndIf
	\end{algorithmic}
\end{algorithm}

Finally, note that $\Delta_\alpha \leq \frac{-2718-1653\alpha-57\alpha^2}{2(72-52\alpha-2\alpha^2)}\alpha$, and $\alpha$ is constant. Lemma~\ref{lemma_approx_alpha} gives us that $A(I) \leq (\sqrt{205}-12+\delta)\opt(I) + \Delta_\alpha$ independently of the value of~$n$. Therefore, if we modify the algorithm to define $n_0(\delta) = 0$ for any~$\delta > 0$, that is, to not solve any instances optimally by enumeration, this modified algorithm has an asymptotic approximation ratio of $\sqrt{205}-12+\delta$ for any~$\delta > 0$.

\section{PTAS}\label{section_ptas}
We now present a PTAS for SMSBPP. The idea of the algorithm is similar to the PTAS for MSBPP presented by \cite{epstein2018}. 

Let~$\ptasvar$ be such that~$1/\varepsilon$ is an integer and ${\ptasvar \leq 1/4}$. We assume~$n \geq 1/\ptasvar^3$. Our algorithm classifies items as small, medium, or large, such that at least~$1/\ptasvar^6$ small items can be packed in the area occupied by a single large item. We also define medium items such that they are few compared to the total number of items in the instance and that their total area is also small compared to the total area of all items.  

We round up the sizes of the large items and find an optimal packing for these items in polynomial time, find an infeasible solution for the small items with value no greater than the value of an optimal solution for these items, and then combine the two packings into a feasible packing without increasing the value of the solution by much. Then, we pack the medium items at the end, which will also not impact the value of the solution much, given how they are chosen.

We now define the medium items. Let $M_i = \{j : s_j \in [\ptasvar^{(3/\ptasvar)(i + 1)}, \ptasvar^{(3/\ptasvar)i})\}$ for $i = 1, \dots, 1/\ptasvar^3$. This defines~$1/\ptasvar^3$ disjoint subsets of the items in an instance, and, therefore, there is at least one value of~$i$ for which $|M_i| \leq \ptasvar^3 n$. For one such value of~$i$, let $M_{i\ell} = \{j \in M_i: s_j \in [\ptasvar^{(3/\ptasvar)i + 3(\ell + 1)}, \ptasvar^{(3/\ptasvar)i + 3\ell})\}$ for\linebreak$\ell = 0, \dots, \frac{1}{\ptasvar}-1$. Again, there is at least one value of~$\ell$ for which\linebreak$\area(M_{i\ell}) \leq \ptasvar \area(I)$. Let $M = M_{i\ell}$, and, for ease of notation, let\linebreak$p =(3/\ptasvar)i+3\ell$. We define the items in~$M$ as medium items, the items larger than the medium items, that is, of size at least~$\ptasvar^p$, as large items (denoted by~$L$), and the items smaller than the medium items, that is, of size less than~$\ptasvar^{p+3}$, as small items (denoted by $S$).

\subsection{Small items}\label{section_ptas_small}

We define a Next Fit Increasing Height (NFIH) heuristic to pack the small items. It is similar to NFDH, however, items are ordered in non-decreasing order of size, and the height of a level is defined as the height of the last item packed in it. When no more levels can be opened in a bin, items are packed above the bin, making the solution infeasible. A new bin is only opened once~$4$ levels have been packed above the current bin. We consider that the cost of items above a bin is still the index of that bin. Let~$Q_j$ be the set of items packed inside bin~$B_j$ and above it. Also, let~$Q$ be the infeasible packing done by NFIH, and~$V(P)$ the cost of a packing~$P$. 

\begin{figure}
\centering
\begin{tikzpicture}
\draw (0,0) rectangle (4,4);

\filldraw[draw=black, fill=lightgray] (0, 0.6-0.1) rectangle (3.9, 3.6-0.1);
\filldraw[draw=black, fill=lightgray] (0, 3.6-0.1) rectangle (3.6, 3.9-0.1);
\draw (0.3, 0.6-0.1) -- (0.3, 3.9-0.1);
\foreach \i in {2, ..., 12}{
	\draw (0, 0.3*\i-0.1) -- (3.9, 0.3*\i-0.1);
	\draw[dashed] (3.9, 0.3*\i-0.1) -- (4, 0.3*\i-0.1);
	\draw (0.3*\i, 0.6-0.1) -- (0.3*\i, 3.9-0.1);
}
\filldraw[draw=black, fill=lightgray] (0,0) rectangle (3.6, 0.15);
\foreach \i in {1, ..., 24}{
	\draw (0.15*\i, 0) -- (0.15*\i, 0.15);
}
\filldraw[draw=black, fill=lightgray] (3.6, 0) rectangle (3.8, 0.2);
\filldraw[draw=black, fill=lightgray] (0, 0.2) rectangle (3.35, 0.41);
\foreach \i in {1, ..., 15}{
	\draw (0.21*\i, 0.2) -- (0.21*\i, 0.41);
}
\filldraw[draw=black, fill=lightgray] (3.35, 0.2) rectangle (3.65, 0.5);
\filldraw[draw=black, fill=lightgray] (3.65, 0.2) rectangle (3.95, 0.5);

\draw (3.8, 0.2) -- (4, 0.2);
\draw[dashed] (3.6, 3.8) -- (4, 3.8);

\draw (-0.2, 4) -- (-0.2, 4.5);
\draw (-0.3, 4) -- (-0.1, 4);
\draw (-0.3, 4.5) -- (-0.1, 4.5);
\draw (-0.5, 4.25) node {$\beta$};

\draw (4.2, 4) -- (4.2, 4.6);
\draw (4.3, 4) -- (4.1, 4);
\draw (4.3, 4.6) -- (4.1, 4.6);
\draw (4.5, 4.3) node {$\beta'$};

\draw (-0.2, 0.2) -- (-0.2, 0.4);
\draw (-0.3, 0.2) -- (-0.1, 0.2);
\draw (-0.3, 0.4) -- (-0.1, 0.4);
\draw (-0.8, 0.3) node {$\geq H_1$};

\draw (4.2, 0.2) -- (4.2, 0.5);
\draw (4.1, 0.2) -- (4.3, 0.2);
\draw (4.1, 0.5) -- (4.3, 0.5);
\draw (4.6, 0.35) node {$H_2$};

\filldraw[draw=black, fill=lightgray] (0, 4) rectangle (3, 4.5);
\foreach \i in {1,...,6}{
	\draw (0.5*\i, 4) -- (0.5*\i, 4.5);
}
\filldraw[draw=black, fill=lightgray] (3, 4) rectangle (3.6, 4.6);
\filldraw[draw=black, fill=lightgray] (0, 4.6) rectangle (0.6*6, 5.2);
\filldraw[draw=black, fill=lightgray] (0, 5.2) rectangle (0.6*6, 5.8);
\filldraw[draw=black, fill=lightgray] (0, 5.8) rectangle (0.7*5, 6.5);
\draw[dashed] (4, 4) -- (4, 6.5);
\draw[dashed] (0, 4.6) -- (4, 4.6);
\draw[dashed] (0, 5.2) -- (4, 5.2);
\draw[dashed] (0, 5.8) -- (4, 5.8);
\draw[dashed] (0, 6.5) -- (4, 6.5);
\draw[dashed] (0, 4) -- (0, 6.5);

\foreach \i in {1,...,6}{
	\draw (0.6*\i, 4.6) -- (0.6*\i, 5.8);
}

\foreach \i in {1,...,5}{
	\draw (0.7*\i, 5.8) -- (0.7*\i, 6.5);
}
\end{tikzpicture}
\caption[Example of a packing done by NFIH, with four levels packed above a bin making the packing infeasible.]{Example of a packing done by NFIH, with four levels packed above a bin, making the packing infeasible. The height~$H_i$ of each level is the size of the last item packed in it, and, thus, the first item packed in the level has size at least~$H_{i-1}$. The items packed above the bin have a size of at least~$\beta$ (size of the first item packed above the bin)}\label{fig_nfih}
\end{figure}
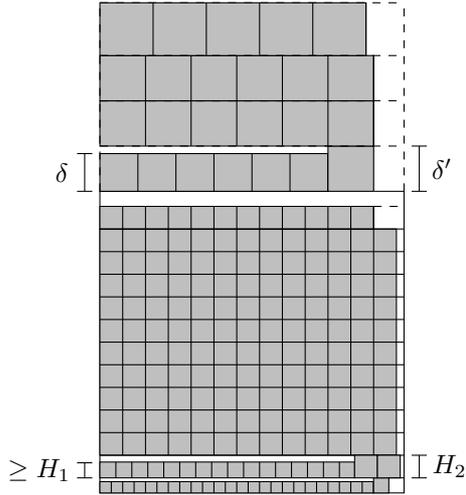

Consider any bin of~$Q$, except the last one. Let~$\beta$ be the size of the first item of the first level packed above the bin, and $\beta' \geq \beta$ be the size of the last item packed in the same level, as shown in Figure~\ref{fig_nfih}. The size of every item inside the bin is at most~$\beta$ and the size of every item above the bin is at least~$\beta$. Let~$H_i$ be the height of the $i$-th level in the bin, with ${H_0 = 0}$. The first item in the $i$-th level has size at least~$H_{i-1}$, and, therefore, the unoccupied area above the items in this level is at most $H_i - H_{i-1}$. Adding this for all levels, we get
\begin{equation*}
\sum_{i=1}^t (H_i - H_{i-1}) = H_t - H_0 = H_t \leq \beta.
\end{equation*}

The empty area above the levels inside the bin is less than~$\beta'$, as, otherwise, the next level would have been packed within the bin. Also, the unoccupied width in every level inside the bin is less than~$\beta$, as, otherwise, another item would have been packed in the level. Therefore, the total unoccupied area inside the bin is less than $2\beta + \beta'$.

Each level above the bin has an occupied width of at least~$1 - \ptasvar^{p+3}$, as, otherwise, another small item would be packed in it. Therefore, the area occupied in the first level above the bin is at least $(1-\ptasvar^{p+3})\beta \geq \frac{3}{4}\beta$, since~$\ptasvar \leq 1/4$. Similarly, the occupied area in the other levels above the bin is at least $(1-\ptasvar^{p+3})\beta' \geq \frac{3}{4}\beta'$. Let~$r$ be the number of bins used. For $1 \leq j < r$,
\begin{equation*}
\area(Q_j) \geq 1 - 2\beta - \beta' + \frac{3}{4}\beta + \frac{9}{4}\beta' \geq 1,
\end{equation*}
and, thus, by Lemma~\ref{lemma_approx_bound}, we have that 
\begin{equation*}
\opt(S) \geq \sum_{j=1}^r j|Q_j| = V(Q).
\end{equation*} 

We now show how to transform~$Q$ into a feasible packing without increasing the value of the solution by much.

\begin{lemma}\label{lemma_ptas_small}
The infeasible packing~$Q$ can be transformed into a feasible packing~$P$ such that $V(P) \leq (1+\ptasvar)V(Q)$.
\end{lemma}
\begin{proof}
Each level above a bin has height less than~$\ptasvar^{p+3}$. Therefore, at least\linebreak$1/\ptasvar^{p+3}$ levels fit in a single bin. We divide the bins of~$Q$ into blocks of\linebreak$\lceil 1/(4\ptasvar^{p+3})\rceil$ bins each (the last block can have fewer bins), and open a new bin in each block to pack the levels above the bins of that block. For $k \geq 2$, the new bin of the $k$-th block is opened at the start of the block, and thus the cost of the items moved to it does not increase. For the first block, the new bin is opened at the $1/\ptasvar$-th position.

The items of the first block in bins of index at least~$1/\ptasvar$ had a cost of at least~$1/\ptasvar$, and this cost increases by~$1$ after the opening of the new bin. Therefore, the increase is by a factor of at most ${1 + \ptasvar}$.

The items of the $k$-th block had a cost of at least~$\frac{k-1}{4\ptasvar^{p+3}}$, and that cost increases by~$k$, since~$k$ new bins are opened before them. Since $\ptasvar \leq 1/4$ and, thus, $p \geq 3/\ptasvar \geq 12$, that increase is by a factor of at most
\begin{equation*}
\left(\frac{k-1}{4\sizeofsmallitem}+k\right)\frac{4\sizeofsmallitem}{k-1} \leq 1 + 8\ptasvar^{15} \leq 1 + \ptasvar.
\end{equation*}

For each bin in which items were packed above the bin, the size of every item inside the bin is at most~$\beta$, while the size of the items above the bin is at least~$\beta$. Therefore, the levels moved to a new bin have at most~$1/\beta$ items each, while the levels that remain inside the bin have at least~$1/\beta$ each. The bin is occupied to at least $1-\sizeofsmallitem$ of its height, as, otherwise, another level with only small items would be packed in it. Thus, at least~$(1-\sizeofsmallitem)/\beta$ levels are packed inside the bin, and the ratio of the number of moved items to the number of items that remain in the bin is at most 
\begin{equation*}
\frac{4}{\beta}\cdot \frac{\beta^2}{1-\sizeofsmallitem} = \frac{4\beta}{1-\sizeofsmallitem} \leq 8\beta \leq 8\sizeofsmallitem < \ptasvar^2.
\end{equation*}

Therefore, the items above the first~$(1/\ptasvar-1)$ bins have the cost increased by a factor of at most~$1/\ptasvar$, and their amount is less than~$\ptasvar^2$ of the total of small items. Thus, the cost of the solution increases by a factor of at most
\begin{equation*}
1+\frac{1}{\ptasvar}\ptasvar^2 = 1 + \ptasvar,
\end{equation*}
and, thus, $V(P) \leq (1+\ptasvar)V(Q)$.\qed
\end{proof}

By Lemmas~\ref{lemma_approx_bound} and~\ref{lemma_ptas_small}, we have that 
\begin{equation*}
V(P) \leq (1+\ptasvar)\opt(S).
\end{equation*}

\subsection{Large items}\label{section_ptas_large}
We now describe how to pack the large items. We do a linear grouping of the large items into~$1/\sizeofgroups$ groups $L_1, \dots, L_{1/\sizeofgroups}$ such that $\lceil \sizeofgroups |L| \rceil = |L_1| \geq |L_2| \geq \cdots \geq |L_{1/\sizeofgroups}| \geq |L_1| - 1$, with~$L_1$ receiving the largest items of the instance,~$L_2$ the next largest items, and so on. After that, we create a new rounded-up instance~$I'$ discarding the items in~$L_1$, and rounding up the sizes of the remaining large items to the size of the largest item in their respective group. With this, the number of distinct item sizes for the large items is bounded by a constant ($1/\sizeofgroups$). Small items remain the same as in~$I$. Since this grouping is very similar to the grouping done in the PTAS for MSBPP by~\cite{epstein2018}, the following lemma still applies.

\begin{lemma}[Epstein et al., 2018]\label{lemma_ptas_large}
A packing~$P'$ of instance~$I'$ can be transformed into a packing~$P$ of instance~$I$ such that $V(P) \leq (1 + 13\ptasvar)V(P')$.
\end{lemma}

Therefore, we focus on finding a packing for~$I'$ first. Let~$L'$ be the large items in~$I'$. Because the number of distinct item sizes in~$L'$ is bounded by a constant, it is possible to enumerate all the possible partitions of items and test if they are a feasible packing in polynomial time (see \cite{2dinapprox}). With this, we can obtain an optimal packing for~$L'$ in polynomial time. Let~$B_{L_1}$ be the first bin of this packing. We now have two different cases for turning this packing into a packing of~$S \cup L'$, depending on the number of small items in the instance.

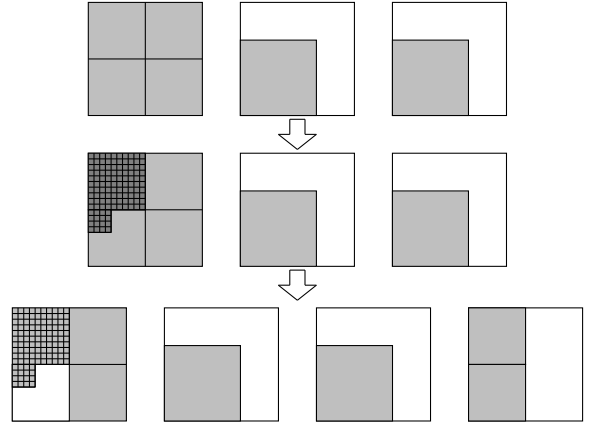
\begin{figure}[H]
\centering
\begin{tikzpicture}[scale=0.5]
\draw[draw=black, fill=lightgray] (0,0) rectangle (3,3);
\draw (4,0) rectangle (7,3);
\draw (8,0) rectangle (11,3);

\draw (1.5, 0) -- (1.5, 3);
\draw (0, 1.5) -- (3, 1.5);
\draw[draw=black, fill=lightgray] (4,0) rectangle (6,2);
\draw[draw=black, fill=lightgray] (8,0) rectangle (10, 2);

\draw (5.3, -0.1) -- (5.3, -0.5); 
\draw (5.7, -0.1) -- (5.7, -0.5); 
\draw (5.3, -0.1) -- (5.7, -0.1);
\draw (5.3, -0.5) -- (5, -0.5);
\draw (5.7, -0.5) -- (6, -0.5);
\draw (5, -0.5) -- (5.5, -0.9);
\draw (6, -0.5) -- (5.5, -0.9);

\draw[draw=black, fill=lightgray] (0,-1) rectangle (3,-4);
\draw (4,-1) rectangle (7,-4);
\draw (8,-1) rectangle (11,-4);
\draw (1.5, -1) -- (1.5, -4);
\draw (0, -2.5) -- (3, -2.5);
\draw[draw=black, fill=lightgray] (4,-2) rectangle (6,-4);
\draw[draw=black, fill=lightgray] (8,-2) rectangle (10, -4);
\draw[draw=black, fill=gray] (0, -1) rectangle (1.5, -2.5);
\draw[draw=black, fill=gray] (0, -2.5) rectangle (0.6, -3.1);

\foreach \i in {1, ..., 10}{
	\draw (0, -1-0.15*\i) -- (1.5, -1-0.15*\i);
	\draw (0.15*\i, -1) -- (0.15*\i, -2.5);
}
\foreach \i in {1, ..., 4}{
	\draw (0, -2.5-0.15*\i) -- (0.6, -2.5-0.15*\i);
	\draw (0.15*\i, -2.5) -- (0.15*\i, -3.1);
}

\draw (5.3, -4.1) -- (5.3, -4.5); 
\draw (5.7, -4.1) -- (5.7, -4.5); 
\draw (5.3, -4.1) -- (5.7, -4.1);
\draw (5.3, -4.5) -- (5, -4.5);
\draw (5.7, -4.5) -- (6, -4.5);
\draw (5, -4.5) -- (5.5, -4.9);
\draw (6, -4.5) -- (5.5, -4.9);
\draw[draw=white] (-2, -5) -- (13, -5);
\end{tikzpicture}
\begin{tikzpicture}[scale=0.5]
\draw[draw=black, fill=lightgray] (0,-5) rectangle (3,-8);
\draw (4,-5) rectangle (7,-8);
\draw (8,-5) rectangle (11,-8);
\draw (1.5, -5) -- (1.5, -8);
\draw (0, -6.5) -- (3, -6.5);
\draw[draw=black, fill=lightgray] (4,-6) rectangle (6,-8);
\draw[draw=black, fill=lightgray] (8,-6) rectangle (10, -8);
\draw[draw=black, fill=white] (0, -6.5) rectangle (1.5, -8);
\draw[draw=black, fill=lightgray] (0, -6.5) rectangle (0.6, -7.1);

\foreach \i in {1, ..., 10}{
	\draw (0, -5-0.15*\i) -- (1.5, -5-0.15*\i);
	\draw (0.15*\i, -5) -- (0.15*\i, -6.5);
}
\foreach \i in {1, ..., 4}{
	\draw (0, -6.5-0.15*\i) -- (0.6, -6.5-0.15*\i);
	\draw (0.15*\i, -6.5) -- (0.15*\i, -7.1);
}

\draw (12, -5) rectangle (15, -8);
\draw[draw=black, fill=lightgray] (12, -5) rectangle (13.5, -6.5);
\draw[draw=black, fill=lightgray] (12, -6.5) rectangle (13.5, -8);
\end{tikzpicture}
\caption[Example of packing when the instance has few small items.]{Example of packing~$P$ when $|S| < |B_{L_1}|/\ptasvar^3$. In the first step, we have an optimal packing of~$L'$. In the next step, small items are packed in the first bin, with the darker region showing the overlap with the large items already packed there. Finally, in the last step, the overlapped large items are moved to a new bin, and the packing becomes feasible}\label{fig_joinspecial}
\end{figure}

Since small items have area of at most~$\areaofsmallitem$ and large items have area of at least~$\areaoflargeitem$, at least
\begin{equation*}
\frac{\areaoflargeitem}{\areaofsmallitem} = \frac{1}{\varepsilon^6}
\end{equation*} 
small items can be packed in the area occupied by a large item by dividing this area into a grid of squares with side~$\sizeofsmallitem$ and packing one small item into each square. Therefore, if $|S| < |B_{L_1}|/\ptasvar^3$, all the small items of the instance can be packed into the areas occupied by $\lceil \ptasvar^3 |B_{L1}| \rceil$ large items packed in~$B_{L_1}$. In this case, we pack the small items in these areas, overlapping some large items, and then open a new bin in the~$1/\ptasvar$-th position, moving the overlapped large items to this new bin, as shown in Figure~\ref{fig_joinspecial}. Let~$P$ be this new packing. 

\begin{lemma}\label{lemma_ptas_joinspecial}
$V(P) \leq (1 + 2\ptasvar)\opt(I)$. 
\end{lemma}
\begin{proof}
As seen in the proof of Lemma~\ref{lemma_ptas_small}, by opening a new bin in the $1/\ptasvar$-th position, items with cost at least~$1/\ptasvar$ have their cost increased by~$1$, and thus the increase is by a factor of at most~$1+\ptasvar$.

The cost of each large item moved from~$B_{L_1}$ to the new bin increases by~$1/\ptasvar-1$. If $\ptasvar^3|B_{L1}| \geq 1$, the total increase in the cost of the solution is at most
\begin{equation*}
\frac{\lceil \ptasvar^3 |B_{L1}| \rceil}{\ptasvar} \leq \frac{2\ptasvar^3|B_{L1}|}{\ptasvar} \leq 2\ptasvar^2|B_{L1}| \leq \ptasvar \opt(I).
\end{equation*}

Otherwise, if $\ptasvar^3|B_{L1}| < 1$, then $|S| < 1/\ptasvar^6$, and thus all the small items are packed overlapping a single large item. Therefore, only one item is moved to the new bin. Since the number of items in the instance is $\opt(I) \geq n \geq 1/\ptasvar^3$, the increase in the cost of this item is at most~$\ptasvar^2 \opt(I)$.\qed
\end{proof}

We now consider the complementary case where $|S| \geq |B_{L_1}|/\ptasvar^3$. In this case, we pack the small items as seen in Lemma~\ref{lemma_ptas_small} and place the bins of the optimal solution for~$L'$ at the end of this packing. Again, let~$P$ be the new packing obtained.

\begin{lemma}\label{lemma_ptas_join}
$V(P) \leq (1+4\ptasvar) \opt(S \cup L')$. 
\end{lemma}
\begin{proof}
Recall that, by Observation~\ref{obs_subsets}, 
\begin{equation*}
\opt(S \cup L') \geq \opt(S) + \opt(L').
\end{equation*} 

By Lemma~\ref{lemma_ptas_small}, we have that the cost of all small items in~$P$ is at most $(1+\ptasvar)\opt(S)$.

Every bin in~$P$ containing small items has at least~$B_S = 1/\areaofsmallitem$ items, except possibly two of them (the last bin of the packing of small items and the last bin opened to repack the items packed in an infeasible way by NFIH), and every bin in~$P$ containing large items has at most~$B_L = 1/\areaoflargeitem$. Thus,
\begin{equation*}
B_L = \frac{1}{\areaoflargeitem} \cdot \areaofsmallitem B_S = \ptasvar^6 B_S,
\end{equation*}
that is, the bins with small items at the start of the solution generally have at least~$1/\ptasvar^6$ times more items than the bins packing large items at the end of the solution.

Let~$m_S$ be the number of bins used to pack the small items in~$P$ and~$m_{L'}$ the number of bins used to pack the items of~$L'$. There are at least $\left\lceil \frac{m_S}{2} \right\rceil$ bins whose indexes are at least $\left\lceil \frac{m_S+1}{2} \right\rceil$. If $m_S \geq 3$, then
\begin{align*}
(1+\ptasvar)\opt(S) &\geq \left(\left\lceil \frac{m_S}{2} \right\rceil - 2\right) \cdot \left\lceil \frac{m_S+1}{2} \right\rceil B_S \\
	&\geq \left(\left\lceil \frac{m_S}{2} \right\rceil - 2\right)\frac{m_S}{2} B_S.
\end{align*}

The cost of every large item increases by~$m_S$ from the optimal solution for~$L'$ to~$P$. For the first $\frac{1}{\ptasvar^3}\left(\left\lceil \frac{m_S}{2} \right\rceil - 2\right)$ bins packing large items, the total increase in the cost of these items is at most
\begin{align*}
\frac{1}{\ptasvar^3}\left(\left\lceil \frac{m_S}{2} \right\rceil - 2\right) m_S B_L &\leq \frac{1}{\ptasvar^3}\left(\left\lceil \frac{m_S}{2} \right\rceil - 2\right) \ptasvar^6 m_S B_S \\
	&\leq 2\ptasvar^3 (1+\ptasvar)\opt(S) \\
	&\leq (2\ptasvar^3 + 2\ptasvar^4)\opt(S) \leq \ptasvar \opt(S).
\end{align*}

The remaining large items have cost at least $\frac{1}{\ptasvar^3}\left(\left\lceil \frac{m_S}{2} \right\rceil - 1\right) \geq \frac{m_S-2}{2\ptasvar^3}$ in the optimal solution for~$L'$, therefore the increase in their cost in~$P$ is by a factor of at most
\begin{equation*}\label{eq_large_items_farther}
1 + \frac{2\ptasvar^3m_S}{m_S-2} \leq 1 + 6\ptasvar^3 \leq 1 + \ptasvar,
\end{equation*}
and thus $V(P)$ is at most 
\begin{equation*}
(1+\ptasvar)\opt(S) + (1+\ptasvar)\opt(L') + \ptasvar \opt(S) \leq (1+2\ptasvar)\opt(S \cup L').
\end{equation*}

Otherwise, if $m_S \leq 3$, then recall that $\opt(S) \geq |S| \geq |B_{L1}|/\ptasvar^3$, and note that~$|B_{L_1}|$ is an upper bound for the number of items of any bin in~$P$ packing large items. For the first~$1/\ptasvar$ bins of the optimal solution for~$L'$, the increase in the cost of the items packed in these bins is at most
\begin{equation*}
\frac{1}{\ptasvar}m_S|B_{L1}| \leq \frac{3}{\ptasvar}\ptasvar^3|S| \leq \frac{3}{\ptasvar}\ptasvar^3\opt(S) = 3\ptasvar^2 \opt(S) \leq \ptasvar \opt(S).
\end{equation*}

The remaining large items have cost of at least~$1/\ptasvar$ in the optimal solution for~$L'$, and so their cost increases by a factor of at most~$1+3\ptasvar$. Thus, $V(P)$ is at most 
\begin{equation*}
(1+\ptasvar)\opt(S) + (1+3\ptasvar) \opt(L') + \ptasvar \opt(S) \leq (1+4\ptasvar)\opt(S \cup L'),
\end{equation*}
and, therefore, in either case $V(P) \leq (1+4\ptasvar) \opt(S \cup L')$.\qed
\end{proof}

By Lemmas~\ref{lemma_ptas_joinspecial} and~\ref{lemma_ptas_join} we have thus that it is possible to pack $S \cup L'$ with cost of at most $(1+4\ptasvar)\opt(I)$.

\subsection{Medium items}\label{section_ptas_medium}
We now show how to pack the medium items. Recall that $|M| \leq \ptasvar^3 n$ and $\area(M) \leq \ptasvar \area(I)$.

\begin{lemma}\label{lemma_ptas_medium}
A packing~$P'$ of $I \backslash M$ can be transformed into a packing~$P$ of~$I$ such that $V(P) \leq (1+7\ptasvar)V(P')$.
\end{lemma}
\begin{proof}
Let~$m$ be the number of bins used by~$P$. We have that
\begin{equation*}
m \geq \area(I \backslash M) \geq (1-\ptasvar)\area(I) \geq \frac{(1-\ptasvar)\area(M)}{\ptasvar} \geq \frac{\area(M)}{2\ptasvar}.
\end{equation*}

We pack the items of~$M$ using NFDH\@. The size of each medium item is at most $\ptasvar^p \leq \ptasvar^{12}$. Therefore, NFDH leaves at most~$2\ptasvar^{12}$ of area unoccupied per bin packing these items, save for the last bin (see \cite{ffdh}). Thus, the number of items needed to pack all medium items is at most
\begin{equation*}
\left\lceil\frac{\area(M)}{1-2\ptasvar^{12}}\right\rceil \leq \lceil2\area(M)\rceil \leq \lceil4\ptasvar m\rceil.
\end{equation*}

For $k = 1, \dots, 1/\ptasvar$, we open four new bins in the $k/\ptasvar$-th position to pack the medium items, as shown in Figure~\ref{fig_ptas_medium}. Every item after the $k/\ptasvar$-th position has cost at least~$k/\ptasvar$ in~$P'$, and~$4k$ bins are opened before them. Therefore, their cost increases by a factor of at most $1+4\ptasvar$.

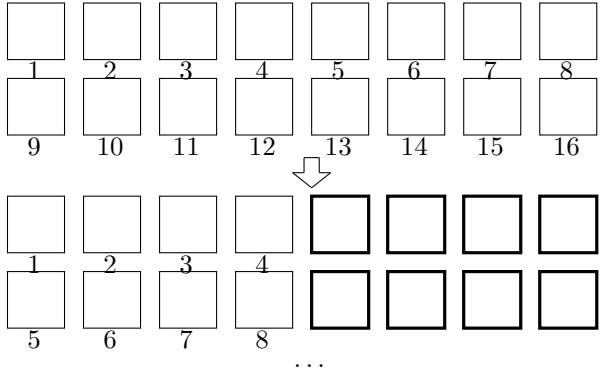
\begin{figure}[H]
\centering
\begin{tikzpicture}[scale=0.5]
\draw (0,-0.5) rectangle (1.5,-2);
\draw (2,-0.5) rectangle (3.5,-2);
\draw (4,-0.5) rectangle (5.5,-2);
\draw (6,-0.5) rectangle (7.5,-2);
\draw (8,-0.5) rectangle (9.5,-2);
\draw (10,-0.5) rectangle (11.5,-2);
\draw (12,-0.5) rectangle (13.5,-2);
\draw (14,-0.5) rectangle (15.5,-2);

\draw (0,0) rectangle (1.5,1.5);
\draw (2,0) rectangle (3.5,1.5);
\draw (4,0) rectangle (5.5,1.5);
\draw (6,0) rectangle (7.5,1.5);
\draw (8,0) rectangle (9.5,1.5);
\draw (10,0) rectangle (11.5,1.5);
\draw (12,0) rectangle (13.5,1.5);
\draw (14,0) rectangle (15.5,1.5);

\foreach \i in {1, ..., 8}{
	\draw (2*\i-2+0.7,-0.3) node{$\i$};
}
\foreach \i in {9, ..., 16}{
	\draw (2*\i-18+0.7,-2.3) node{$\i$};
}
\draw (7.8, -2.6) -- (7.8, -3); 
\draw (8.2, -2.6) -- (8.2, -3); 
\draw (7.8, -2.6) -- (8.2, -2.6);
\draw (7.8, -3) -- (7.5, -3);
\draw (8.2, -3) -- (8.5, -3);
\draw (7.5, -3) -- (8, -3.4);
\draw (8.5, -3) -- (8, -3.4);

\draw[draw=white] (0, -3.5) -- (1, -3.5);
\end{tikzpicture}
\begin{tikzpicture}[scale=0.5]
\draw (0,-0.5) rectangle (1.5,-2);
\draw (2,-0.5) rectangle (3.5,-2);
\draw (4,-0.5) rectangle (5.5,-2);
\draw (6,-0.5) rectangle (7.5,-2);
\draw[very thick] (8,-0.5) rectangle (9.5,-2);
\draw[very thick] (10,-0.5) rectangle (11.5,-2);
\draw[very thick] (12,-0.5) rectangle (13.5,-2);
\draw[very thick] (14,-0.5) rectangle (15.5,-2);

\draw (0,0) rectangle (1.5,1.5);
\draw (2,0) rectangle (3.5,1.5);
\draw (4,0) rectangle (5.5,1.5);
\draw (6,0) rectangle (7.5,1.5);
\draw[very thick] (8,0) rectangle (9.5,1.5);
\draw[very thick] (10,0) rectangle (11.5,1.5);
\draw[very thick] (12,0) rectangle (13.5,1.5);
\draw[very thick] (14,0) rectangle (15.5,1.5);

\foreach \i in {1, ..., 4}{
	\draw (2*\i-2+0.7,-0.3) node{$\i$};
}
\foreach \i in {5, ..., 8}{
	\draw (2*\i-10+0.7,-2.3) node{$\i$};
}
\draw (8, -3) node{$\dots$};
\end{tikzpicture}
\caption[Example of packing with $\ptasvar = 1/4$.]{Example of bins of a packing with $\ptasvar = 1/4$. The bins with thick borders in the second step are the bins opened to pack the medium items}\label{fig_ptas_medium}
\end{figure}

Since we opened~$4/\ptasvar$ bins to pack the medium items, the cost of packing each of them is at most~$1/\ptasvar^2 + 4/\ptasvar$. Recall that the number of medium items is $|M| \leq \ptasvar^3 n$ and that $\ptasvar \leq 1/4$. The total cost of packing all medium items is at most
\begin{equation*}
\left(\frac{1}{\ptasvar^2}+\frac{4}{\ptasvar}\right)\ptasvar^3 n = (\ptasvar + 4\ptasvar^2) n \leq 2\ptasvar n = \frac{2\ptasvar}{1-\ptasvar^3} (1-\ptasvar^3)n \leq 3(1-\ptasvar^3)n,
\end{equation*}
and if $|M| \leq \ptasvar^3 n$, then $V(P') \geq |I \backslash M| \geq (1-\ptasvar^3)n$, and so
\begin{equation*}
V(P) \leq (1+4\ptasvar)V(P') + 3\ptasvar V(P') \leq (1+7\ptasvar) V(P'),
\end{equation*}
thus the result follows.\qed
\end{proof}

\subsection{The PTAS algorithm}\label{section_ptas_final}
We can now present the full algorithm for our PTAS\@.

\begin{algorithm}\label{alg_ptas}
	\caption{PTAS for SMSBPP}
	\begin{algorithmic}
		\Require {List $I$ of $n$ square items, $\varepsilon > 0$}
		\State Step 1: Classify the items of the instance into small, medium, or large, as described in the beginning of Section~\ref{section_ptas}.
		\State Step 2: Create a rounded-up instance~$I'$ as described in Section~\ref{section_ptas_large}, and find an optimal solution~$P_{L'}$ for~$L'$ (the rounded up large items).
		\State Step 3: Transform~$P_{L'}$ into a packing~$P'$ for $S \cup L'$ in one of the ways described in Section~\ref{section_ptas_large}, depending on the number of small items in the instance.
		\State Step 4: Convert~$P'$ into a packing~$P$ for~$I \backslash M$ by opening~$|L_1|$ new bins to pack the large items discarded in the creation of~$I'$.
		\State Step 5: Finally, pack the medium items as described in Section~\ref{section_ptas_medium}.
	\end{algorithmic}
\end{algorithm}

With the lemmas discussed throughout the section, we now show that this algorithm is a PTAS for the SMSBPP\@.

\begin{theorem}\label{thm_ptas}
The algorithm proposed in this section is a PTAS for SMSBPP\@.
\end{theorem}

\begin{proof}
After transforming $P_{L'}$ into a packing~$P'$ for $S \cup L'$ as described in Step 3, we have, by Lemmas~\ref{lemma_ptas_joinspecial} and~\ref{lemma_ptas_join}, that 
\begin{equation*}
	V(P') \leq (1+4\ptasvar)\opt(I).
	\end{equation*}

Converting~$P'$ into a packing~$P$ in Step 4, we have that, by Lemma~\ref{lemma_ptas_large},
\begin{align*}
V(P) &\leq (1+13\ptasvar)V(P') \\
	&\leq (1+13\ptasvar)(1+4\ptasvar)\opt(I)\\ 
	&\leq (1+17\ptasvar+52\ptasvar^2)\opt(I) \\
	&\leq (1+30\ptasvar)\opt(I).
\end{align*}

Finally, after packing the medium items in Step 5, by Lemma~\ref{lemma_ptas_medium}, the final packing has value of at most
\begin{align*}
(1+7\ptasvar)(1+30\ptasvar)\opt(I)&= (1+37\ptasvar+210\ptasvar^2)\opt(I) \\
&\leq (1+90\ptasvar)\opt(I).
\end{align*}

Therefore, for some $\alpha \leq \ptasvar/90$, the algorithm finds solutions of value at most $(1+\alpha)\opt(I)$ in polynomial time. Thus, the algorithm is a PTAS for SMSBPP\@.\qed
\end{proof}

\section{Conclusion}\label{section_concl}
In this work, we investigated the {\squareminsum}, presenting an approximation algorithm of simple implementation and $O(n \log n)$ time complexity, which also has an absolute approximation ratio for instances with only small items. We have also presented a PTAS for the problem. The approximation ratio of~$\sqrt{205}-12+\delta$ that we have obtained for the first algorithm is not proven to be tight, and a better analysis might be possible for future work. One possible direction for that is finding better lower bounds for the value of an optimal solution.


%
\section*{Statements and Declarations}

\paragraph{Conflict of interest} The authors declare that they have no conflict of interest.

\paragraph{Data availability} This manuscript has no associated data.

\sloppypar
\paragraph{Funding} Supported by Grants 404779/2025-5, 312345/2023-2, and 163644/2021-7, National Council for Scientific and Technological Development (CNPq). This study was financed in part by the Coordenação de Aperfeiçoamento de Pessoal de Nível Superior - Brasil (CAPES) - Finance Code 001. Supported by Grant 2022/05803-3, São Paulo Research Foundation (FAPESP). Supported by Grant 2372/23, Fundo de Apoio ao Ensino, Pesquisa e Extensão - FAEPEX/Unicamp.

\section*{Acknowledgement}
This version of the article has been accepted for publication, after peer review (when applicable) but is not the Version of Record and does not reflect post-acceptance improvements, or any corrections. The Version of Record is
available online at: https://doi.org/10.1007/s10878-026-01425-4.

\bibliographystyle{spbasic}      
\bibliography{article_bib}   

\end{document}